\documentclass[%
reprint,
superscriptaddress,
amsmath,amssymb,
prl,
]{revtex4-1}

\usepackage{graphicx}% Include figure files
\usepackage{dcolumn}% Align table columns on decimal point
\usepackage{bm}% bold math
\usepackage{natbib}
\usepackage[mathlines]{lineno}% Enable numbering of text and display math
\usepackage{color}

\begin{document}
%\preprint{APS/123-QED}

%\title{First direct, precision determination of the atomic mass of a superheavy element}
% -or-
%\title{Determination of the atomic number of a superheavy nuclide by direct, precision atomic mass measurement}% isotope}% Force line breaks with \\
% -or-
\title{First high-precision direct determination of the atomic mass of a superheavy nuclide}% evinces a new means to unambiguously determine the atomic numbers of superheavy elements}
%\title{First high-precision direct determination of the atomic mass of $^{257}$Db evinces a means to unambiguously determine atomic number}


\author{P~Schury}
\email{schury@post.kek.jp}
\affiliation{KEK Wako Nuclear Science Center, Wako, Saitama 351-0198, Japan}

\author{T. Niwase}
\affiliation{Department of Physics, Kyushu University, Nishi-ku, Fukuoka 819-0395, Japan}
\affiliation{RIKEN Nishina Center for Accelerator-Based Science, Wako, Saitama 351-0198, Japan}
\affiliation{KEK Wako Nuclear Science Center, Wako, Saitama 351-0198, Japan}

\author{M.~Wada}
\affiliation{KEK Wako Nuclear Science Center, Wako, Saitama 351-0198, Japan}

\author{P.~Brionnet}
\affiliation{RIKEN Nishina Center for Accelerator-Based Science, Wako, Saitama 351-0198, Japan}

\author{S.~Chen}
\affiliation{%
Department of Physics, The University of Hong Kong, Pokfulam, 999077, Hong Kong
}%
\affiliation{KEK Wako Nuclear Science Center, Wako, Saitama 351-0198, Japan}

\author{T.~Hashimoto}
\affiliation{%
Institute for Basic Science, 70, Yuseong-daero 1689-gil, Yusung-gu, Daejeon, Korea
}%

\author{H.~Haba}
\affiliation{RIKEN Nishina Center for Accelerator-Based Science, Wako, Saitama 351-0198, Japan}

\author{Y.~Hirayama}
\affiliation{KEK Wako Nuclear Science Center, Wako, Saitama 351-0198, Japan}

\author{D.S.~Hou}
\affiliation{Institute of Modern Physics, Chinese Academy of Sciences, Lanzhou 730000, China}
\affiliation{University of Chinese Academy of Sciences, Beijing 100049, China}
\affiliation{School of Nuclear Science and Technology, Lanzhou University,  Lanzhou 730000, China}

\author{S.~Iimura}
\affiliation{Department of Physics, Osaka University, Osaka, Japan}
\affiliation{RIKEN Nishina Center for Accelerator-Based Science, Wako, Saitama 351-0198, Japan}
\affiliation{KEK Wako Nuclear Science Center, Wako, Saitama 351-0198, Japan}

\author{H.~Ishiyama}
\affiliation{RIKEN Nishina Center for Accelerator-Based Science, Wako, Saitama 351-0198, Japan}

\author{S.~Ishizawa}
\affiliation{Graduate School of Science and Engineering, Yamagata University, Yamagata, Japan}
\affiliation{RIKEN Nishina Center for Accelerator-Based Science, Wako, Saitama 351-0198, Japan}

\author{Y.~Ito}
\affiliation{Advanced Science Research Center, Japan Atomic Energy Agency (JAEA), Tokai, Ibaraki 319-1195, Japan}
\affiliation{RIKEN Nishina Center for Accelerator-Based Science, Wako, Saitama 351-0198, Japan}
\affiliation{KEK Wako Nuclear Science Center, Wako, Saitama 351-0198, Japan}

\author{D.~Kaji}
\affiliation{RIKEN Nishina Center for Accelerator-Based Science, Wako, Saitama 351-0198, Japan}

\author{S.~Kimura}
\affiliation{RIKEN Nishina Center for Accelerator-Based Science, Wako, Saitama 351-0198, Japan}

\author{H.~Koura}
\affiliation{Advanced Science Research Center, Japan Atomic Energy Agency (JAEA), Tokai, Ibaraki 319-1195, Japan}

\author{J.J.~Liu}
\affiliation{%
Department of Physics, The University of Hong Kong, Pokfulam, 999077, Hong Kong
}%
\affiliation{KEK Wako Nuclear Science Center, Wako, Saitama 351-0198, Japan}

\author{H.~Miyatake}
\affiliation{KEK Wako Nuclear Science Center, Wako, Saitama 351-0198, Japan}

\author{J.-Y.~Moon}
\affiliation{%
Institute for Basic Science, 70, Yuseong-daero 1689-gil, Yusung-gu, Daejeon, Korea
}%

\author{K.~Morimoto}
\affiliation{RIKEN Nishina Center for Accelerator-Based Science, Wako, Saitama 351-0198, Japan}

\author{K.~Morita}
\affiliation{Department of Physics, Kyushu University, Nishi-ku, Fukuoka 819-0395, Japan}
\affiliation{Research Center for SuperHeavy Elements, Kyushu University, Nishi-ku, Fukuoka 819-0395, Japan}

\author{D.~Nagae}
\affiliation{Research Center for SuperHeavy Elements, Kyushu University, Nishi-ku, Fukuoka 819-0395, Japan}

\author{M.~Rosenbusch}
\affiliation{KEK Wako Nuclear Science Center, Wako, Saitama 351-0198, Japan}

\author{A.~Takamine}
\affiliation{RIKEN Nishina Center for Accelerator-Based Science, Wako, Saitama 351-0198, Japan}

\author{Y.X.~Watanabe}
\affiliation{KEK Wako Nuclear Science Center, Wako, Saitama 351-0198, Japan}

\author{H.~Wollnik}
\affiliation{New Mexico State University, Las Cruces, NM 88001, USA}

\author{W.~Xian}
\affiliation{%
Department of Physics, The University of Hong Kong, Pokfulam, 999077, Hong Kong
}%
\affiliation{KEK Wako Nuclear Science Center, Wako, Saitama 351-0198, Japan}

\author{S.X.~Yan}
\affiliation{Institute of Mass Spectrometer and Atmospheric Environment, Jinan University, Guangzhou, 510632, China}


\date{\today}% It is always \today, today,

\begin{abstract}
%The RIKEN Ring Cyclotron at the Nishina Center for Accelerator-Based Science was used to produce a
We present the first direct measurement of the atomic mass of a superheavy nuclide.  Atoms of $^{257}$Db ($Z$=105) were produced online at the RIKEN Nishina Center for Accelerator-Based Science using the fusion-evaporation reaction $^{208}$Pb($^{51}$V, 2n)$^{257}$Db.  The gas-filled recoil ion separator GARIS-II was used to suppress both the unreacted primary beam and some transfer products, prior to delivering the energetic beam of $^{257}$Db ions to a helium gas-filled ion stopping cell wherein they were thermalized.  Thermalized $^{257}$Db$^{3+}$ ions were then transferred to a multi-reflection time-of-flight mass spectrograph for mass analysis.  An alpha particle detector embedded in the ion time-of-flight detector allowed disambiguation of the rare $^{257}$Db$^{3+}$ time-of-flight detection events from background by means of correlation with characteristic $\alpha$-decays. The extreme sensitivity of this technique allowed a precision atomic mass determination from 11 events.  The mass excess was determined to be $100\,063(231)_\textrm{stat}(132)_\textrm{sys}$~keV/c$^2$.  Comparing to several mass models, we show the technique can be used to unambiguously determine the atomic number as $Z$=105 and should allow similar evaluations for heavier species in future work.
%\begin{description}
%\end{description}
\end{abstract}

\pacs{Valid PACS appear here}% PACS, the Physics and Astronomy
\maketitle

%\tableofcontents

%Introduction

\par The unambiguous identification of superheavy nuclei is a longstanding issue that has largely been achieved through cross-bombardment experiments in recent years \cite{Oganessian2013, Oganessian2004, MoritaZ113}.  For the hot fusion superheavy nuclides (SHN) located ``northeast'' of $^{263}$Rf, however, cross-bombardment does not fully resolve the question \cite{Forsberg2016} as all such nuclides thus far produced exhibit decay chains which terminate in spontaneous fission prior to reaching well-known nuclides.  Results from efforts to unambiguously determine $Z$ using characteristic X-rays \cite{x-ray} have yet to garner widespread acceptance, either.   The Provisional Report of the 2017 Joint Working Group of IUPAC and IUPAP \cite{JWP2017} suggested that direct determination of the atomic mass with a sufficient precision could, in many cases, be a valid means to fully determine the $A$ and $Z$ of an uncertain nuclide, particularly if decay information were simultaneously obtained.  A first effort in this direction has recently shown some promise by directly verifying the mass number of a superheavy nuclide \cite{FIONA2018}, however without a level of mass precision needed to confirm the atomic number.

\par Beyond identification of SHN, the precise determination of atomic masses is vital to understanding the heaviest elements.  Proper evaluation of the possible production -- both in the laboratory \cite{Production_Sobiczewski, Production_Oganessian} and in the cosmos \cite{CosmicSHE_Petermann, CosmicSHE_Flerov} -- of nuclides in the island of stability, theorized to be composed of exceptionally long-lived SHN \cite{Wheeler56}, requires accurate atomic masses in the heavy and superheavy region.

\par Among isotopes of transuranium elements, directly determined atomic masses are rare \cite{SHIPTRAP_No, TRIGATRAP, Ito2018}; for SHN ($Z \ge 104$) directly determined atomic masses are completely absent.  Atomic masses of most $Z>92$ nuclides are determined indirectly, often through long decay chains.  Herein we report the first direct measurement of the atomic mass of an SHN, $^{257}$Db. The measurement was performed by combination of a multi-reflection time-of-flight mass spectrograph (MRTOF-MS) and the newly developed ``$\alpha$-TOF" detector \cite{Niwase2019}, greatly improving the sensitivity of the MRTOF-MS by providing correlational data between time-of-flight (ToF) and subsequent $\alpha$-decay.  This measurement, the first to utilize $\alpha$-decay correlated time-of-flight mass spectroscopy, serves as both a cross-check for our previous indirect measurement of the mass of $^{257}$Db \cite{Ito2018} and a proof-of-principle for future efforts to measure nuclides in the hot-fusion superheavy island which do not connect to well-known nuclei via $\alpha$-decay, and where typical yields will be on the order of a few per day.

%Apparatus and measurement technique
\par Atoms of $^{257}$Db were produced in the fusion-evaporation reaction $^{208}$Pb($^{51}$V, 2n)$^{257}$Db \cite{Gates2008}.  The RIKEN Ring Cyclotron provided a 306~MeV beam of $^{51}$V$^{13+}$ with maximum intensity of $\approx$500~pnA.  The beam impinged upon a rotating target wheel comprised of aluminum energy degraders and $^{208}$Pb targets.  The targets were made from $^{208}$Pb enriched to 99.6\% and deposited on a 30~$\mu$g/cm$^2$ carbon backing, with a typical lead thickness of 360~$\mu$g/cm$^2$.  Aluminum energy degraders of 12~$\mu$m thickness were utilized to reduce the beam energy to 243~MeV at target center.  A detector angled 45$^\circ$ to the beam axis, located near the target wheel, measured the rate of elastic recoils from the target, providing a means to measure the effective primary beam dose.

\begin{figure}[t]
	\vspace{2mm}
	\centering
	\includegraphics[width = 3.4in]{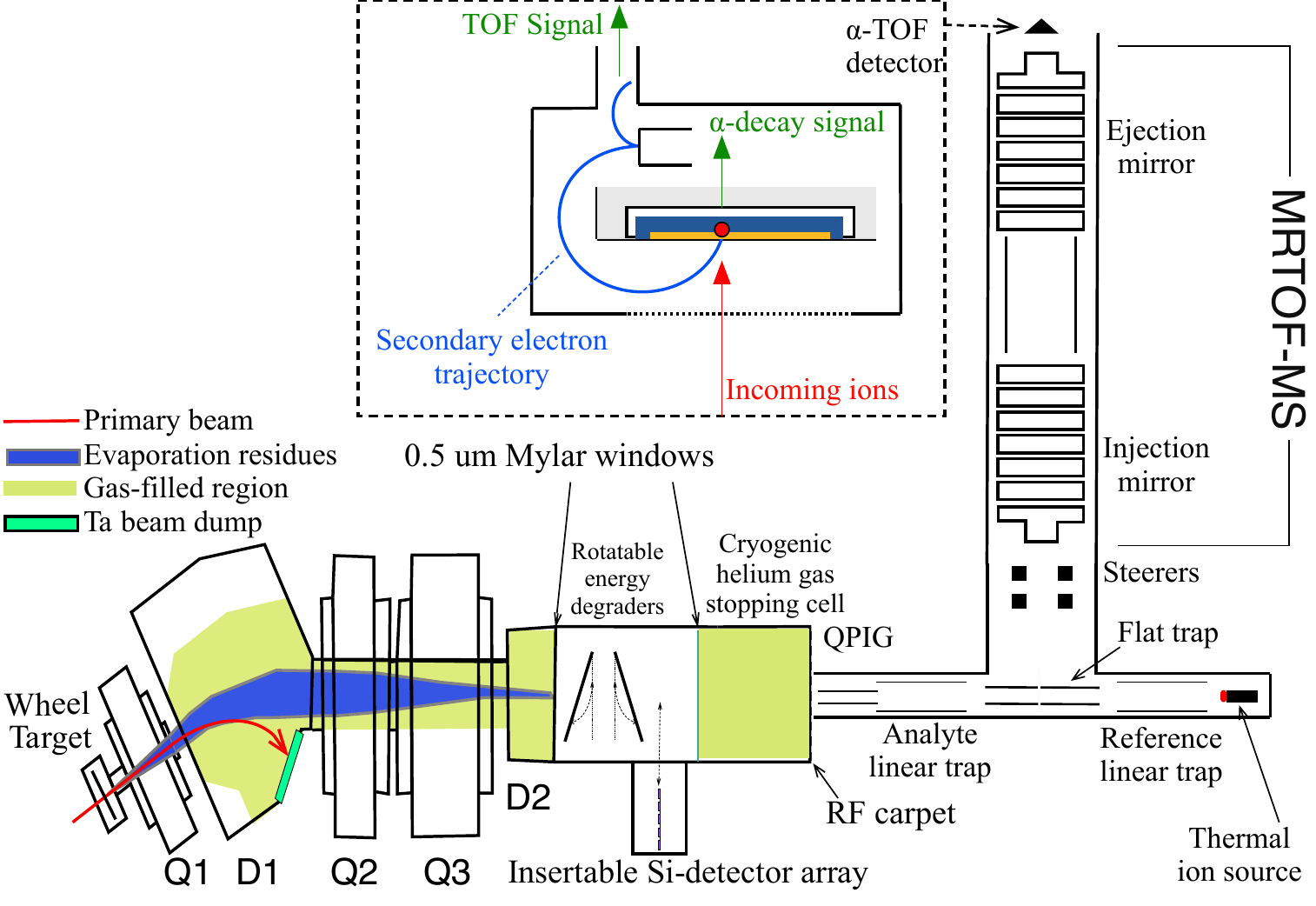}
	\caption{\label{figApparatus} Sketch (not to scale) of the apparatus used in the measurement.  Dubnium atoms are produced via fusion-evaporation reactions.  Ions of the fusion-evaporation products are separated from the primary beam using the gas-filled recoil ion separator GARIS-II.  The ions are stopped in the helium gas cell and subsequently stored in an RF ion trap before being sent to a multi-reflection time-of-flight mass spectrograph (MRTOF-MS) for analysis.  The ion detector at the end of the MRTOF-MS can detect ion implantation and subsequent $\alpha$-decay.    \vspace{-1.5 mm}} 
\end{figure}

\par The gas-filled recoil ion separator GARIS-II \cite{GARIS-II} transported $^{257}$Db while suppressing the primary beam and various transfer products.  It was filled with dilute helium gas at 70~Pa. From previous experience with $^{257}$Db \cite{Brionnet257Db} the selective dipole (D1 in Fig.~\ref{figApparatus}) was set to 1.42~T.

\par As shown in Fig.~\ref{figApparatus}, after exiting GARIS-II the beam passed through rotatable Mylar energy degraders prior to entering a helium-filled gas cell.  The gas cell was cryogenically cooled to 60~K and pressurized to 200~mbar room temperature equivalent.  The Mylar degrader thickness was chosen to reduce the energy of $^{257}$Db to be commensurate with the stopping power of the helium in the gas cell.  A static electric field transported stopped ions to a traveling-wave radio-frequency (RF) ion carpet \cite{Wada2003, Arai2014} with a 0.74~mm diameter exit orifice.  After exiting the gas cell, ions were transported through a differentially pumped region by use of quadrupole RF ion guides and trapped in a segmented linear Paul trap, which is part of a three-trap suite used to prepare analyte and reference ions for analysis by the MRTOF-MS using the concomitant referencing method \cite{Ito2018,SchuryWideband2} that allows analyte ions to be accumulated with nearly 100\% duty cycle.  The traps were cryogenically cooled to $\approx$150~K to minimize the probability of stored ions charge exchanging with residual background gases.
 
\par The MRTOF-MS, a device finding widespread use in recent years \cite{Knauer2019, Fischer2018, Wienholtz2013, Wolf2013a, PLA2013457, Leistenschneider2018, HIRSH2016229, SCHULTZ2016251, CHAUVEAU2016, WANG2020179, Murray2019}, consists of a pair of ion mirrors separated by a field-free drift region.  The outermost electrode of each mirror is switched to allow ions to enter and exit.  Ions are stored in the MRTOF-MS for a time sufficient to allow the ions to reflect a specific number of times and achieve a time focus.  During the measurement reported herein, the mass resolving power at the time focus was typically $R_m$$\approx$250\,000, with flight times of $t$$\sim$10~ms for $A/q\approx85$ ions.  To preclude detector dead time leading to undercounting that could affect the reference peak shape, the reference ion source was adjusted so as to detect one reference ion ($^{85}$Rb$^+$ or $^{133}$Cs$^+$) per cycle on average. 

\par Stable molecular ions produced in the gas cell or transfer products not removed by GARIS-II may have mass-to-charge ratios significantly differing from the analyte ion and will make fewer or more reflections than the analyte ions and may, by happenstance, appear at the same ToF as the analyte ions.  As such, erroneous attribution is a concern with MRTOF-MS measurements \cite{SchuryWideband2, Fischer2018}.  In the case of analyte ions detected at a rate of a few per day, however, confidence in the ability to exclude background noise (dark counts\cite{DCR}, cosmic rays, and $\alpha$- or $\beta$-decay from {\emph e.g.} transfer product ions), or even extremely low-yield molecular ions with mass-to-charge ratio nearly identical to the analyte, becomes an issue of concern. 

\par To overcome these issues we have developed a novel ``$\alpha$-TOF" detector \cite{Niwase2019} based on a commercial MagneToF ion detector.  Incoming ions strike a specially coated impact plate, which then releases secondary electrons.  The secondary electrons are isochronously guided by a permanent magnet through an electron multiplier to produce a detectable ion impact signal.  An ion's time-of-flight, defined as the duration starting with ejection from the ion trap and ending with detection of the ion impact signal, is measured using a time-to-digital converter (MCS6 from FAST ComTec).
\par We have embedded a silicon PIN diode in the impact plate of a MagneToF ion detector.  The PIN diode's energy resolution is $\sigma_\textrm{E}$$\approx$140~keV.  High-confidence measurements can be achieved by evaluating  ``$\alpha$-decay-correlated ToF events" in which an $\alpha$-decay event (``$\alpha$ single") of a proper energy is observed within a proper duration subsequent to an ion impact signal (``ToF single") with timing consistent with the expected analyte ion.

\par Since the $\alpha$-TOF detector's location precludes $\alpha$-particle energy calibration by offline sources, $^{185}$Hg was produced via the $^{139}$La($^{51}$V, 5n) reaction prior to production of $^{257}$Db.  The 5653~keV and 5372~keV $\alpha$-particles from the $\alpha$-decay of $^{185}$Hg \cite{1976GrZC,NNDC185Hg} were used to calibrate the $\alpha$-TOF's silicon PIN diode.

\par Separately, the incoming rate of $^{185}$Hg was measured on an insertable silicon PIN diode array located between GARIS-II and the gas cell.  Using the measured rate of $^{185}$Hg$^{2+}$ in MRTOF-MS time-of-flight spectra, the efficiency from gas cell through to $\alpha$-TOF was determined to be between 4\% and 5\% for ToF detection. 

%Analysis technique
\par Ions impact the $\alpha$-TOF detector with an energy of $\approx$2~keV/$q$, implanting a few angstroms deep and geometrically limiting the detection efficiency to 45\%. The recoil from an $\alpha$-particle emitted toward the detector is sufficient to eject the atom from the detector surface. Thus sequential $\alpha$-particles along the decay chain cannot be observed.  
Fortunately, when an $\alpha$ particle is emitted away from the detector, the daughter is generally not removed from the surface and there is a similar 45\% probability for detection of the daughter's $\alpha$-decay.  The lifetimes of nuclides in the $^{257}$Db decay chain allow the evaluation to extend out four decays, through $^{245}$Es.  Accounting for the $\alpha$-decay branching ratios of each nuclide \cite{Hessberger2001, Briselet2019} the total likelihood to detect one of the $\alpha$-decays in the $^{257}$Db decay chain would be 65\%.

\par An initial effort to measure $^{257}$Db$^{2+}$ (based on previous experience \cite{SchuryDoubleCharge,Rosenbusch2018,Ito2018}) produced no correlated event candidates after a dose on target of 4.7$\times$10$^{17}$ particles.  As the National Institute of Standards and Technology (NIST) Atomic Spectra Database \cite{NIST} lists the 3$^\textrm{rd}$ ionization potential of Db as being 23.1$\pm$1.6~eV, compared to helium's 24.6~eV 1$^{st}$ ionization potential, it was possible that $^{257}$Db$^{3+}$ ions would be delivered from the gas cell.  As such, the transport conditions were set for $A/q$$\sim$85 and an effort was made to measure $^{257}$Db$^{3+}$.  During 105~hours of measurement, with a total dose of 1.1$\times$10$^{18}$~particles, a total of 14 decay-correlated event candidates were observed, which was consistent with the evaluated system efficiency.  

\par
\begin{figure}[t]
%	\vspace{2mm}
	\centering
	 \includegraphics[width = 3.4in]{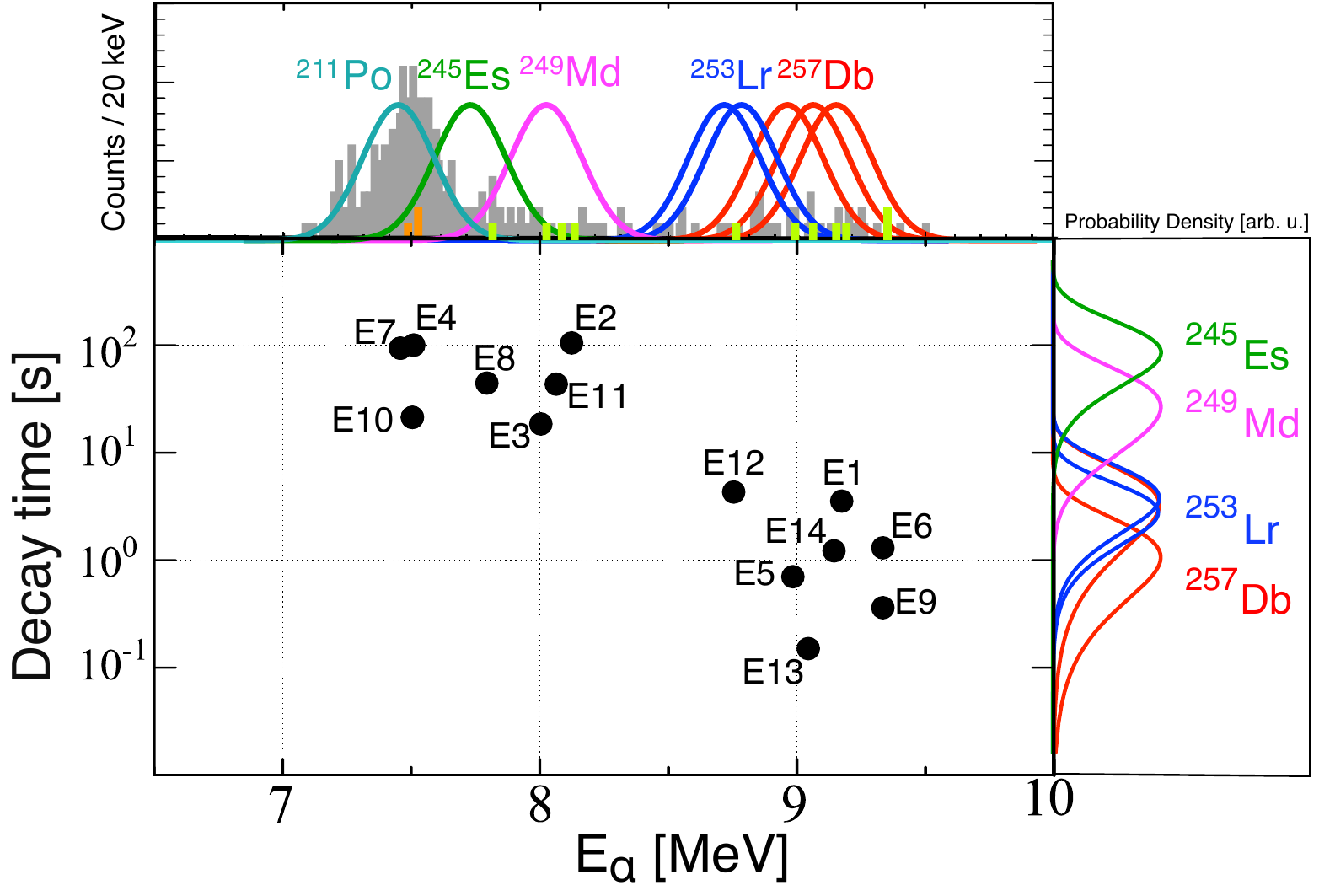} 
	 \vspace{-6.5 mm}
	\caption{\label{figEvents}   Distribution of ToF-correlated $\alpha$-decay events in terms of detected $\alpha$-decay energy ($E_\alpha$) and decay time.  For nuclides in the decay chain of $^{257}$Db, the probability distributions in terms of decay time \cite{Schmidt1984} are shown at right; two curves are given for $^{257}$Db and $^{253}$Lr as they each exhibit isomerism.  The detector response function for each $\alpha$-decay is shown at top, overlaying the $\alpha$ singles spectrum with correlated event candidates denoted by colored marks. The multiple $\alpha$-decay channels which exist for $^{253}$Lr and $^{257}$Db are shown. 
\vspace{-2.5 mm}} 
\end{figure}

 A ToF gate such that $t\in t_c\pm50$~ns, where $t_c$ is the expected ToF of $^{257}$Db$^{3+}$  based on the 2016 Atomic Mass Evaluation (AME16) \cite{AME16} excludes superfluous ions.  An energy gate of $E_\alpha\ge$7.0~MeV encompass all $\alpha$-decays from $^{245}$Es, $^{249}$Md, $^{253}$Lr, and $^{257}$Db (see top panel of Fig.~\ref{figEvents}).  Any gated ToF single followed within 120~s by a gated $\alpha$-decay single was considered a correlated event candidate.

\par Figure \ref{figEvents} plots the correlated events observed in this work in terms of detected $\alpha$-decay energy and decay time; events are named in order of occurrence.  The right panel of Fig.~\ref{figEvents} shows the anticipated decay time probability distributions \cite{Schmidt1984} for each nuclide; multiple curves are shown for $^{257}$Db and $^{253}$Lr to represent known isomers.  Similarly, the upper panel shows the detector response curves for each $\alpha$-decay which could be observed in the $^{257}$Db decay chain.  The events form two clusters corresponding to $^{257}$Db or $^{253}$Lr and $^{249}$Md or $^{245}$Es. 

\par
\begin{figure}[b]
%	\vspace{2mm}
	\centering
	 \includegraphics[width = 3.4in]{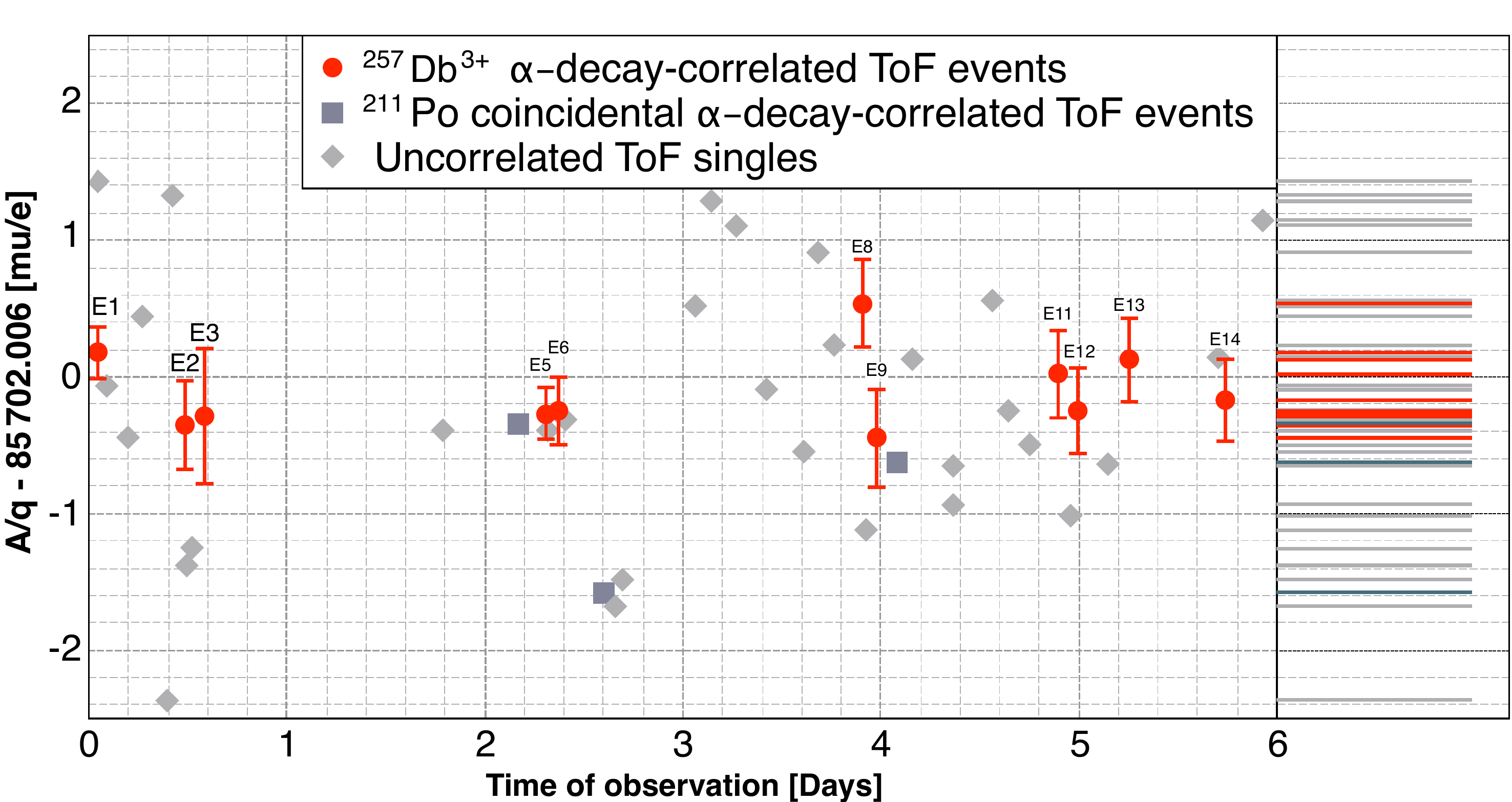} 
	 \vspace{-5mm}
	\caption{\label{figSingles} Apparent $A/q$ evaluated for each ToF single near the expected position of $^{257}$Db$^{3+}$.  The data are plotted in terms of deviation from the $A/q$ for $^{257}$Db$^{3+}$ as determined from AME16.  Statistical uncertainties are only evaluated for $\alpha$-decay correlated ToF events.  
\vspace{-5 mm}}
\end{figure}

 The full spectrum (above 7~MeV) of $\alpha$ singles is shown in grey in the top panel of Fig.~\ref{figEvents}.  The large peak centered near 7.5~MeV is presumed to be from $^{211}$Po.  While the $N$=126 neutron shell closure in $^{208}$Pb may suppress direct production of $^{211g}$Po ($T_{1/2}$=516~ms) at this beam energy, its $EC$ decay parent $^{211}$At ($T_{1/2}$=7.214~h) can be produced in the decay of extremely short-lived multi-nucleon transfer products $^{215}$Fr ($T_{1/2}$=86~ns), $^{219}$Ac ($T_{1/2}$=12~$\mu$s), and $^{223}$Pa ($T_{1/2}$=5.1~ms).  While evidence of $^{211m,g}$Po$^{2+}$ was identified in the time-of-flight spectra, its short-lived progenitors were not seen, likely having decayed before being extracted from the gas cell.  Based on the 120~s coincidence window and 235 observed $\alpha$-decays commensurate with $^{211}$Po in the course of 105~hours of data accumulation, during which 37 ToF singles events in the vicinity of $^{257}$Db$^{3+}$ were observed (see Fig.~\ref{figSingles}), we could expect to observe $\approx$3 coincidental correlations with $^{211}$Po decay.  A similar evaluation \cite{NiwaseThesis} indicates that less than one coincidental correlation would be expected for the higher-energy $\alpha$-decay signals.  In consideration of this, we exclude events E4, E7, and E10 from our analysis of the atomic mass of $^{257}$Db as their energies and decay times are more consistent with $^{211}$Po than with any nuclide in the $^{257}$Db decay chain.

\par To determine the atomic mass, we typically make use of a single-reference method \cite{ItoSRM} to evaluate the mass of an analyte ion using only one species of reference ion.  The mass-to-charge ratio of the analyte ion can then be related to that of the reference ion by $(A/q)_\textrm{analyte} = \rho^2\cdot(A/q)_\textrm{reference}$.  The value $\rho^2$ is the actual experimental data, given by 
$\rho^2 = \big{(}\frac{t_\textrm{analyte} - t_0}{t_\textrm{reference} - t_0}\big{)}^2$, where $t_0$ 
represents some inherent delay between the ions starting their movement in the analyzer and the start of the clock, while $t_\textrm{analyte}$ and $t_\textrm{reference}$ are the times-of-flight of the analyte and reference, respectively. Based on $\rho^2$( $^{85}$Rb$^+$ / $^{208}$Pb$^{2+}$) measured at the end of the online experiment, it was determined that $t_0$=75(4)~ns. 

\par To exclude confusing an ion with significantly different $A/q$ for our intended analyte ion, spectra are typically made at different numbers of oscillations in the MRTOF-MS reflection chamber \cite{SchuryWideband2}.  The times-of-flight $t_\textrm{analyte}$ and $t_\textrm{reference}$ would typically be determined by fitting the analyte and reference ions' spectra with a response function known to well-reproduce the data.  In this work, however, it was not possible to perform such fittings on the analyte ions' spectral peaks as the number of events at any given number of laps did not exceed three.  Rather, for each analyte ion we made such a fitting for the reference ions spanning 7.5~s before and after the analyte ion's detection and used the individual analyte ion's ToF as $t_\textrm{analyte}$.  The result of this analysis is shown in Fig.~\ref{figSingles}, where the $\alpha$-decay-correlated ToF events are distinguished from uncorrelated events.  The data for the $\alpha$-decay-correlated ToF events are tabulated in Table~\ref{tabResults} along with the various characterizing qualities of each.

\par As the data is an admixture of spectra having differing flight paths, reliable fitting is precluded and we use algebraic weighted averaging to deduce $A/q$ of $^{257}$Db$^{3+}$.  The spectral peak is known to exhibit a slight asymmetry which could lead to a systematic error in such an evaluation.  To ascertain the likely degree of such error, we used a data set comprising 3\,358 consecutive sets of 10 analyte ions of $^{185}$Au$^{2+}$ taken during preparation for the $^{257}$Db measurements.  Mulltiplying the uncertainty of each data set by its Birge ratio \cite{Birge} produced a nearly normal distribution:  51\% were within 1-$\sigma$ of the AME16-derived $\rho^2$-value, 83\% within 2-$\sigma$, 95\% within 3-$\sigma$.

\begin{table}[t]
\vspace{-2mm}
\caption{Summary of correlated ToF-$\alpha$ events, showing the number of times the $^{257}$Db$^{3+}$ reflects back-and-forth in the MRTOF-MS (laps), the observed $\alpha$-particle energy ($E_\alpha$) and the time between implantation and decay ($\Delta t_\alpha$), and the best estimate as to the nuclide which emitted the detected $\alpha$-particle in each correlated event (Nuclide).  The $\rho^2$ column provides the evaluated ratio of mass-to-charge ratios for $^{257}$Db$^{3+}$ and $^{85}$Rb$^+$, see text for details.}
\vspace{-1mm}
\begin{center}
\begin{tabular}{|c|c|c|r|c|l|}
\hline
Event & laps & $E_\alpha$ [MeV] & $\Delta t_\alpha$  [s] & Nuclide & \multicolumn{1}{c|}{ $\rho^2$}  \\
\hline
E1 & 300 &9.19 & 3.54 & $^{257}$Db  & 1.009\,314\,964(90)\\
E2 & 300 &8.14 & 105.00 & $^{249}$Md  & 1.009\,308\,647(157)\\
E3 & 300 &8.02 & 18.50 & $^{249}$Md  & 1.009\,309\,454(237)\\
%E4 & 325 &7.52 & 100.20 & $^{211}$Po  & 1.009308879(158)\\
E5 & 325 &9.00 & 0.70 & $^{257}$Db  & 1.009\,309\,712(91)\\
E6 & 325 &9.35 & 1.30 & $^{257}$Db  & 1.009\,309\,926(119)\\
%E7 & 325 &7.48 & 93.9 & $^{211}$Po  & 1.009294298(150)\\
E8 & 324 &7.81 & 44.00 & $^{245}$Es  & 1.009\,319\,206(155)\\
E9 & 324 &9.35 & 0.36 & $^{257}$Db  & 1.009\,307\,610(173)\\
%E10 & 324 &7.52 & 21.20 & $^{211}$Po & 1.009305524(111)\\
E11 & 327 &8.08 & 43.40 & $^{249}$Md  & 1.009\,309\,949(156)\\
E12 & 327 &8.77 & 4.30 & $^{253}$Lr  & 1.009\,313\,092(150)\\
E13 & 331 &9.06 & 0.15 & $^{257}$Db  & 1.009\,314\,345(148)\\
E14 & 331 &9.16 & 1.20 & $^{257}$Db  & 1.009\,310\,844(144)\\
\hline
\multicolumn{5}{|c|}{Weighted Average} & 1.009\,311\,901(40)\\
\multicolumn{5}{|c|}{ AME16-derived Value} & 1.009\,312\,860(840)\\
\hline
\multicolumn{5}{|c|}{Birge ratio} &  \multicolumn{1}{c|}{24.4} \\
\multicolumn{5}{|c|}{Reweighted Average} & 1.009\,311\,901(973)\\
\hline

\end{tabular}
\vspace{-7mm}
\end{center}
\label{tabResults}
\end{table}%

\par  After renormalizing the uncertainties, the weighted average ratio of $\frac{A/q(^{257}\textrm{Db}^{3+})}{A/q(^{85}\textrm{Rb}^+)}$ was determined to be $\rho^2$=1.009\,311\,901(973)$_\textrm{stat}$(7)$_\textrm{sys}$, the systematic uncertainty deriving from $\delta t_0$=4~ns.  This gives a mass excess of $100\,063(231)$$_\textrm{stat}$(2)$_\textrm{sys}$~keV/c$^2$.  The directly determined value differs from the previous value, which was determined indirectly from $Q_\alpha$-values connecting $^{257}$Db to $^{249}$Md, by $-171(321)$~keV/c$^2$.  This indicates that the accepted $Q_\alpha$-values for $^{257}$Db and $^{253}$Lr are accurate on the 100~keV level.
%, a 171(231)~keV/c$^2$ reduction in binding energy compared to AME16.

\par $^{257}$Db exhibits at least one long-lived isomeric state \cite{Hessberger2001}.  Neither the state order nor the isomeric excitation have been confirmed as yet.  While NUBASE \cite{NUBASE2016} presently recommends an isomeric excitation of 140~keV based on systematics, $\alpha$-decay studies of $^{257}$Db populated by $\alpha$-decay of $^{261}$Bh suggest a 370~keV isomeric excitation \cite{Hessberger2010}.  In this work, the mass resolution of our MRTOF nor the energy resolution of the $\alpha$-TOF detector were sufficient to resolve the two states in $^{257}$Db.  As such, we must supplement the statistical uncertainty in the measured atomic mass with a systematic uncertainty accounting for the admixture of ground and isomer.  A recent study of $^{257}$Db \cite{PierreThesis} indicates an isomeric yield of 39(7)\% for the shorter-lived state.  As such, we assume the ToF-correlated $\alpha$-decay events measured were nearly evenly split between the two states.  Splitting the difference between NUBASE and Ref.~\cite{Hessberger2010} we therefore add 130~keV/c$^2$ systematic uncertainty.  

%Conclusion 
\par
\begin{figure}[t]
%	\vspace{2mm}
	\centering
	 \includegraphics[width = 3.4in]{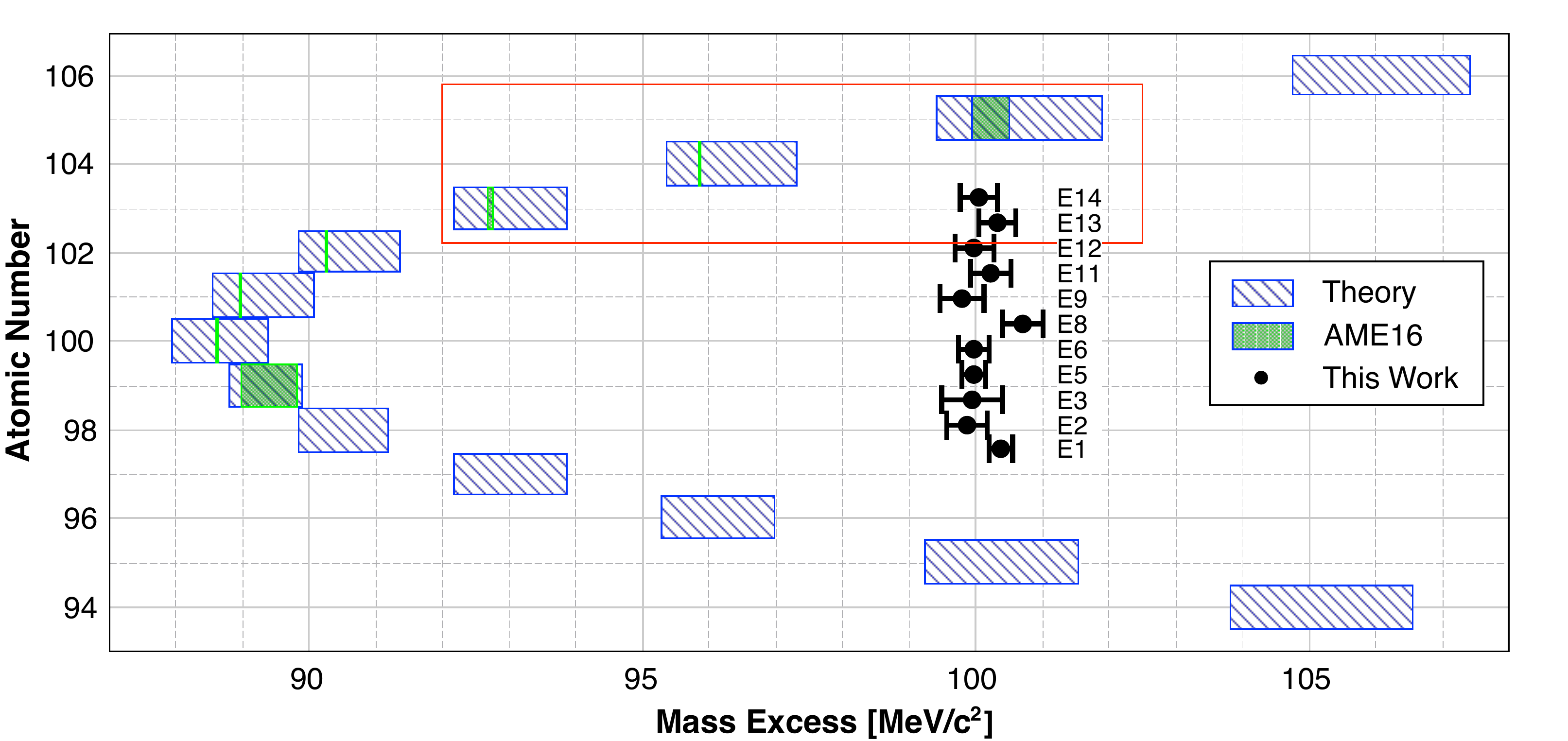} 
	\caption{\label{figDetermineZ} Mass excess determined for each $\alpha$-decay correlated ToF event in this work compared to mass excess ranges for $A$=257 isobars as determined by various mass models \cite{DZ, DZ1, FRDM12, WS4RBF, HFB21, HFB32, WB03, KTUY05} (blue hash) along with values from AME16  \cite{AME16} (green hash).  The red box designates nuclides whose production would be physically possible in the $^{208}$Pb($^{51}$V, X) reaction.
\vspace{-8.0 mm}} 
\end{figure}

 Among SHN, identification becomes ever more challenging with distance from species which could be produced in macroscopic quantities, particularly for production by hot fusion -- where charge particle evaporation is more likely than in cold fusion -- and in multinucleon transfer reactions.  To demonstrate the degree to which the present technique may be applied to such a problem, consider Fig.~\ref{figDetermineZ}.  We present the span of $A$=257 mass excess predictions from a comprehensive selection of global mass models \cite{DZ1, DZ, FRDM12, WS4RBF, HFB21, HFB32, WB03, KTUY05} (in blue), along with those from AME16 (in green).  We superimpose on that the mass excess measured for each $\alpha$-decay-correlated ToF event presumed to correspond to $^{257}$Db$^{3+}$.  

\par In the $^{208}$Pb($^{51}$V, 2n)$^{257}$Db reaction, the $A$=257 nuclides which could be produced are limited by available neutrons and protons in the compound nucleus, as shown by the red box in Fig.~\ref{figDetermineZ}.  As such, it is clear that the analyte ion was $^{257}$Db.  It is worth noting that, had a multi-nucleon transfer reaction been employed, the $\alpha$-TOF detector could use differences in the decay properties of $^{257}$Db and $^{257}$Am to distinguish between them.

\par Considering the demonstrated measurement precision and the typical variance of mass excess from theoretical models, it would be feasible to precisely determine $Z$ in many -- but certainly not all -- cases of presently-known nuclides by $\alpha$-decay correlated ToF spectroscopy.

\par In this letter we have presented a new technique to mass analyze extremely low-yield species using $\alpha$-decay correlated ToF spectroscopy, and demonstrated a method of mass evaluation based on single-ions.  Over the course of a five day online measurement we observed eleven $^{257}$Db correlated $\alpha$-ToF events, from which the mass excess of $^{257}$Db was determined to be $100\,063(231)$$_\textrm{stat}$(132)$_\textrm{sys}$~keV/c$^2$ ($\delta m/m_\textrm{stat}$=9.7$\times$10$^{-7}$), in good agreement with our previous indirect mass determination \cite{Ito2018}, and a direct determination of $Z$ by comparison with mass models was demonstrated.   Additionally, the observed predominance of triply-charged $^{257}$Db delivered from the helium gas cell indicates that the third ionization potential of dubnium must be less than 24.5~eV, restricting the range given by NIST \cite{NIST}.  

\par  The techniques presented here will be used in future measurements to directly confirm the identities of hot-fusion superheavy nuclides sufficiently far from the valley of $\beta$-decay stability, such as $^{288}$Mc/Fl, having sufficient separation to discern $Z$.  Eventually, it may be applied to identification of multi-nucleon transfer products.  Such reactions populate both sides of the valley of stability, making identification by mass spectroscopy impossible without utilizing decay correlated measurement. 

\par To better resolve isomeric states in future measurements, efforts are underway to improve the energy resolution of the $\alpha$-TOF detector.  Similarly, improvements in the MRTOF mass resolving power will allow the precision presented herein to be achieved with as few as 3 correlated $\alpha$-ToF events in future measurements; if the isomer in $^{257}$Db has an excitation energy exceeding 300~keV it could be resolved in the ToF spectrum.

\par We wish to express gratitude to the Nishina Center for Accelerator-Based Science at RIKEN and the Center for Nuclear Study at the University of Tokyo for their support of online measurements.  This work was supported by the Japan Society for the Promotion of Science KAKENHI (Grant Numbers 2200823, 24224008, 24740142, 15H02096, 17H06090, 19K03899, 18H03711, and 15K05116).

%\par Something about future usefulness with MNT studies?


\begin{thebibliography}{9}

\bibitem{Oganessian2013} Yu. Ts. Oganessian, F. Sh. Abdullin, S. N. Dmitriev, J. M. Gostic, J. H. Hamilton, R. A. Henderson, M. G. Itkis, K. J. Moody, A. N. Polyakov, A. V. Ramayya, J. B. Roberto, K. P. Rykaczewski, R. N. Sagaidak, D. A. Shaughnessy, I. V. Shirokovsky, M. A. Stoyer, N. J. Stoyer, V. G. Subbotin, A. M. Sukhov, Yu. S. Tsyganov, V. K. Utyonkov, A. A. Voinov, and G. K. Vostokin, Physical Review C 87 (2013) 014302

\bibitem{Oganessian2004} Yu. Ts. Oganessian, V. K. Utyonkov, Yu. V. Lobanov, F. Sh. Abdullin, A. N. Polyakov, I. V. Shirokovsky, Yu. S. Tsyganov,G. G. Gulbekian, S. L. Bogomolov, B. N. Gikal, A. N. Mezentsev, S. Iliev, V. G. Subbotin, A. M. Sukhov, A. A. Voinov,G. V. Buklanov, K. Subotic, V. I. Zagrebaev, and M. G. Itkis, J. B. Patin, K. J. Moody, J. F. Wild, M. A. Stoyer, N. J. Stoyer, D. A. Shaughnessy, J. M. Kenneally, and R. W. Lougheed, Physical Review C 69 (2004) 054607

\bibitem{MoritaZ113} Kosuke Morita, Kouji Morimoto, Daiya Kaji, Hiromitsu Haba, Kazutaka Ozeki, Yuki Kudou, Takayuki Sumita, Yasuo Wakabayashi, Akira Yoneda, Kengo Tanaka, Sayaka Yamaki, Ryutaro Sakai, Takahiro Akiyama, Shin-ichi Goto, Hiroo Hasebe, Minghui Huang, Tianheng Huang, Eiji Ideguchi, Yoshitaka Kasamatsu, Kenji Katori, Yoshiki Kariya, Hidetoshi Kikunaga, Hiroyuki Koura, Hisaaki Kudo, Akihiro Mashiko, Keita Mayama, Shin-ichi Mitsuoka, Toru Moriya, Masashi Murakami, Hirohumi Murayama, Saori Namai, Akira Ozawa, Nozomi Sato, Keisuke Sueki, Mirei Takeyama, Fuyuki Tokanai, Takayuki Yamaguchi, and Atsushi Yoshida, J. Phys. Soc. Jpn.81, 103201 (2012)

\bibitem{Forsberg2016}  U. Forsberg, D. Rudolph, C. Fahlander, P. Golubev, L.G. Sarmiento, S. {\AA}berg, M. Block, Ch.E. D\"ullmann, F.P. He{\ss}berger, J.V. Kratz, A. Yakushev, Physics Letters B 760 (2016) 293-296.

\bibitem{x-ray} D. Rudolph, U. Forsberg, P. Golubev, L. G. Sarmiento, A. Yakushev, L.-L. Andersson, A. Di Nitto, Ch. E. D\"ullmann, J. M. Gates, K. E. Gregorich, C. J. Gross, F. P. He{\ss}berger, R.-D. Herzberg, J. Khuyagbaatar, J. V. Kratz, K. Rykaczewski, M. Sch\"adel, S.{\AA}berg, D. Ackermann, M. Block, H. Brand, B. G. Carlsson, D. Cox, X. Derkx, K. Eberhardt, J. Even, C. Fahlander, J. Gerl, E. J\"ager, B. Kindler, J. Krier, I. Kojouharov, N. Kurz, B. Lommel, A. Mistry, C. Mokry, H. Nitsche, J. P. Omtvedt, P. Papadakis, I. Ragnarsson, J. Runke, H. Schaffner, B. Schausten, P. Th\"orle-Pospiech, T. Torres, T. Traut, N. Trautmann, A. T\"urler, A. Ward, D. E. Ward, and N. Wiehl, Physical Review Letters 111 (2013) 112502

\bibitem{JWP2017} Sigurd Hofmann, Sergey N. Dmitriev, Claes Fahlander, Jacklyn M. Gates, James B. Roberto and Hideyuki Sakai, Pure and Applied Chemistry 90 (2018) 1774--1832 

\bibitem{FIONA2018} J.M. Gates, G.K. Pang, J.L. Pore, K.E. Gregorich, J.T. Kwarsick, G. Savard, N.E. Esker, M. Kireeff Covo, M.J. Mogannam, J.C. Batchelder, D.L. Bleuel, R.M. Clark, H.L. Crawford, P. Fallon, K.K. Hubbard, A.M. Hurst, I.T. Kolaja, A.O. Macchiavelli, C. Morse, R. Orford, L. Phair, and M.A. Stoyer, Physical Review Letters 121 (2018) 222501

\bibitem{Production_Sobiczewski}A. Sobiczewski and K. Pomorski, Prog. Part. Nucl. Phys 58 (2007) 292

\bibitem{Production_Oganessian}  Y. T. Oganessian, V. K. Utyonkov, Y. V. Lobanov, F. S. Abdullin, A. N. Polyakov, R. N. Sagaidak, I. V. Shirokovsky, Y. S. Tsyganov, A. A. Voinov, G. G. Gulbekian, S. L. Bogomolov, B. N. Gikal, A. N. Mezentsev, S. Iliev, V. G. Subbotin, A. M. Sukhov, K. Subotic, V. I. Zagrebaev, G. K. Vostokin, M. G. Itkis, K. J. Moody, J. B. Patin, D. A. Shaughnessy, M. A. Stoyer, N. J. Stoyer, P. A. Wilk, J. M. Kenneally, J. H. Landrum, J. F. Wild, and R. W. Lougheed, Physical Review C 74 (2006) 044602

\bibitem{CosmicSHE_Petermann} I. Petermann, K. Langanke, G. Mart\'{\i}nez-Pinedo, I. V. Panov, P.-G. Reinhard, F.-K. Thielemann, Eur. Phys. J. A 48 (2012) 122

\bibitem{CosmicSHE_Flerov}  G.N. Flerov, G.M. Ter-Akopian, Rep. Prog. Phys. 46 (1983) 817

\bibitem{Wheeler56} J.A. Wheeler, Proc. Int. Conf. Peaceful Uses of Atomic Energy 2, 155, 220 (1956) (United Nations, New York)

%\bibitem{FissionRecyclingShibagaki} S. Shibagaki, T.Kajino, G. J. Mathews, S. Chiba, S. Nishimura, G. Lorusso, The Astrophysical Journal, Volume 816, Number 2 (2016)

%\bibitem{Julin2001} R. Julin, Nuclear Physics A 685, 221 (2001), nucleusNucleus Collisions 2000

%\bibitem{Bucurescu2013} D. Bucurescu and N. V. Zamfir, Physical Review C 87 (2013) 054324

%\bibitem{Block2010} M. Block, D. Ackermann, K. Blaum, C. Droese, M. Dworschak, S. Eliseev, T. Fleckenstein, E. Haettner, F. Herfurth, F. P. He{\ss}berger, S. Hofmann, J. Ketelaer, J. Ketter, H.-J. Kluge, G. Marx, M. Mazzocco, Y. N. Novikov, W. R. Pla{\ss}, A. Popeko, S. Rahaman, D. Rodr\'{i}guez, C. Scheidenberger, L. Schweikhard, P. G. Thirolf, G. K. Vorobyev, and C. Weber, Nature 463 (2010) 785

\bibitem{Ito2018} Y. Ito, P. Schury, M. Wada, F. Arai, H. Haba, Y. Hirayama, S. Ishizawa, D. Kaji, S. Kimura, H. Koura, M. MacCormick, H. Miyatake, J.Y. Moon, K. Morimoto, K. Morita, M. Mukai, I. Murray, T. Niwase, K. Okada, A. Ozawa, M. Rosenbusch, A. Takamine, T. Tanaka, Y.X. Watanabe, H. Wollnik, and S. Yamaki, Physical Review Letters 120 (2018) 152501

\bibitem{SHIPTRAP_No}M. Dworschak, M. Block, D. Ackermann, G. Audi, K. Blaum, C. Droese, S. Eliseev, T. Fleckenstein, E. Haettner, F. Herfurth, F. P. He{\ss}berger1, S. Hofmann, J. Ketelaer, J. Ketter, H.-J. Kluge, G. Marx, M. Mazzocco, Yu. N. Novikov, W. R. Pla{\ss}, A. Popeko, S. Rahaman, D. Rodr\'{i}guez, C. Scheidenberger, L. Schweikhard, P. G. Thirolf, G. K. Vorobyev, M. Wang, and C. Weber, Physical Review C 81 (2010) 064312

\bibitem{TRIGATRAP}M. Eibach, T. Beyer, K. Blaum, M. Block, Ch. E. D\"ullmann,, K. Eberhardt, J. Grund, Sz. Nagy, H. Nitsche, W. N\"ortersh\"auser, D. Renisch, K. P. Rykaczewski, F. Schneider, C. Smorra, J. Vieten, M. Wang, and K. Wendt, Physical Review C 89 (2014) 064318 

\bibitem{Niwase2019} T. Niwase, M. Wada, P. Schury, H. Haba, S. Ishizawa, Y. Ito, D. Kaji, S. Kimura, H. Miyatake, K. Morimoto, K. Morita, M. Rosenbusch, H. Wollnik, T. Shanley, Y. Benari, Nuclear Instruments and Methods in Physics Research, Section A: Accelerators, Spectrometers, Detectors and Associated Equipment, 953 (2019) 163198, DOI:10.1016/j.nima.2019.163198

\bibitem{Gates2008} J. M. Gates, S. L. Nelson, K. E. Gregorich, I. Dragojevi{\'c}, Ch. E. D{\"u}llmann, P. A. Ellison, C. M. Folden III, M. A. Garcia, L. Stavsetra, R. Sudowe, D. C. Hoffman, and H. Nitsche Phys. Rev. C 78 (2008) 034604

\bibitem{GARIS-II} D. Kaji, K. Morimoto, N. Sato, A. Yoneda, K. Morita, Nuclear Instruments and Methods in Physics Research Section B: Beam Interactions with Materials and Atoms  317B (2013) 311-314

\bibitem{Brionnet257Db} Private Communication with P. Brionnet

\bibitem{Wada2003} Michiharu Wada, Yoshihisa Ishida, Takashi Nakamura, Yasunori Yamazaki, Tadashi Kambara, Hitoshi Ohyama, Yasushi Kanai, Takao M. Kojima, Youichi Nakai, Nagayasu Ohshima, Atsushi Yoshida, Toshiyuki Kubo, Yukari Matsuo, Yoshimitsu Fukuyama, Kunihiro Okada, Tetsu Sonoda, Shunsuke Ohtani, Koji Noda, Hirokane Kawakami, and Ichiro Katayama, Nuclear Instruments and Methods in Physics Research Section B: Beam Interactions with Materials and Atoms 204 (2003) 570

\bibitem{Arai2014} F. Arai, Y. Ito, M. Wada, P. Schury, T. Sonoda, and H. Mita, Int. J. Mass Spectrom. 362 (2014) 56-58

\bibitem{SchuryWideband2} P. Schury, Y. Ito, M. Rosenbusch, H. Miyatake, M. Wada, H. Wollnik, Int. J. Mass Spectrom. 433 (2018) 40-46


%\bibitem{Isomerism_LiuWalker} H. L. Liu, P. M. Walker, and F. R. Xu, Physical Review C 89 (2014) 044304

%\bibitem{Isomerism_Herzberg}Herzberg, R., Greenlees, P., Butler, P. et al., Nature 442 (2006) 896-899

%\bibitem{Isomerism_Khuyagbaatar}J. Khuyagbaatar, A.K. Mistry, D. Ackermann, L.L. Andersson, M. Block, H. Brand, Ch.E. D\"ullmann, J. Even, F.P. He{\ss}berger, J. Hoffmann, A. H\"ubner, E. J\"ager, B. Kindler, J. Krier, N. Kurz, B. Lommel, B. Schausten, J. Steiner, A. Yakushev, V. Yakusheva, Nuclear Physics A (2020) 121662

%\bibitem{isomerism_Dracoulis} G. D. Dracoulis, P. M. Walker, and F. G. Kondev, Rep. Prog. Phys. 79, 076301 (2016).

\bibitem{Knauer2019} S. Knauer, P. Fischer, G. Marx, M. M\"{u}ller, M. Rosenbusch, B. Schabinger, L. Schweikhard, R.N. Wolf, International Journal of Mass Spectrometry 446 (2019) 116189

\bibitem{Fischer2018} Paul Fischer, Gerrit Marx, Lutz Schweickhard, International Journal of Mass Spectrometry 435 (2019) 305-314

\bibitem{DCR} Y. Kang, H. X. Lu, and Y.-H. Lo, Appl. Phys. Lett. 83, 2955 (2003)

\bibitem{Wienholtz2013} F. Wienholtz, D. Beck, K. Blaum, Ch. Borgmann, M. Breitenfeldt, R.B. Cakirli, S. George, F. Herfurth, J.D. Holt, M. Kowalska, S. Kreim, D. Lunney, V. Manea, J. Men\`endez, D. Neidherr, M. Rosenbusch, L. Schweikhard, A. Schwenk, J. Simonis, J. Stanja, R.N. Wolf,  and K. Zuber, Nature 498 (2013) 346-349

\bibitem{Wolf2013a} R.N. Wolf, F. Wienholtz, D. Atanasov, D. Beck, K. Blaum, Ch. Borgmann, F. Herfurth, M. Kowalska, S. Kreim, Yu. A. Litvinov, D. Lunney, V. Manea, D. Neidherr, M. Rosenbusch, L. Schweikhard, J. Stanja, and K. Zube, International Journal of Mass Spectrometry 349-350 (2013) 349-350

\bibitem{PLA2013457} W.R. Pla{\ss}, T. Dickel, S. Purushothaman, P. Dendooven, H. Geissel, J. Ebert, E. Haettner, C. Jesch, M. Ranjan, M.P. Reiter, H. Weick, F. Amjad, S. Ayet, M. Diwisch, A. Estrade, F. Farinon, F. Greiner, N. Kalantar-Nayestanaki, R. Kn\"{o}bel, J. Kurcewicz, J. Lang, I. Moore, I. Mukha, C. Nociforo, M. Petrick, M. Pf\"{u}tzner, S. Pietri, A. Prochazka, A.-K. Rink, S. Rinta-Antila, D. Sch\"{a}fer, C. Scheidenberger, M. Takechi, Y.K. Tanaka, J.S. Winfield, and M.I. Yavor, Nuclear Instruments and Methods in Physics Research Section B: Beam Interactions with Materials and Atoms 317 (2013) 457-462

\bibitem{Leistenschneider2018} E. Leistenschneider, M.P. Reiter, S. Ayet San Andr\'es, B. Kootte, J.D. Holt, P. Navr\'atil, C. Babcock, C. Barbieri, B.R. Barquest, J. Bergmann, J. Bollig, T. Brunner, E. Dunling, A. Finlay, H. Geissel, L. Graham, F. Greiner, H. Hergert, C. Hornung, C. Jesch, R. Klawitter, Y. Lan, D. Lascar, K.G. Leach, W. Lippert, J.E. McKay, S.F. Paul, A. Schwenk, D. Short, J. Simonis, V. Som\`a, R. Steinbr\"ugge, S.R. Stroberg, R. Thompson, M.E. Wieser, C. Will, M. Yavor, C. Andreoiu, T. Dickel, I. Dillmann, G. Gwinner, W.R. Pla\ss{}, C. Scheidenberger, A.A. Kwiatkowski, J. and Dilling, Physical Review Letters 120 (2018) 062503

\bibitem{HIRSH2016229} Tsviki Y. Hirsh, Nancy Paul, Mary Burkey, Ani Aprahamian, Fritz Buchinger, Shane Caldwell, Jason A. Clark, Anthony F. Levand, Lin Ling Ying, Scott T. Marley, Graeme E. Morgan, Andrew Nystrom, Rodney Orford, Adrian P\'{e}rez Galv\'{a}n, John Rohrer, Guy Savard and Kumar S. Sharma, Kevin Siegl, Nuclear Instruments and Methods in Physics Research Section B: Beam Interactions with Materials and Atoms 376 (2016) 229-232

\bibitem{SCHULTZ2016251} B.E. Schultz, J.M. Kelly, C. Nicoloff, J. Long, S. Ryan, and M. Brodeur, Nuclear Instruments and Methods in Physics Research Section B: Beam Interactions with Materials and Atoms 376 (2016) 251-255

\bibitem{CHAUVEAU2016} P. Chauveau, P. Delahaye, G. De France, S. El Abir, J. Lory, Y. Merrer, M. Rosenbusch, L. Schweikhard, and R.N. Wolf, Nuclear Instruments and Methods in Physics Research Section B: Beam Interactions with Materials and Atoms 376 (2016) 211-215

\bibitem{WANG2020179} Jun-Ying Wang, Yu-Lin Tian, Yong-Sheng Wang, Zai-Guo Gan, Xiao-Hong Zhou, Hu-Shan Xu, and Wen-Xue Huang, Nuclear Instruments and Methods in Physics Research Section B: Beam Interactions with Materials and Atoms 463 (2020) 179-183

\bibitem{Murray2019} K. Murray, J. Dilling, R. Gornea, Y. Ito, T. Koffas, A.A. Kwiatkowski, Y. Lan, M.P. Reiter, V. Varentsov, T. Brunner, and the nEXO collaboration, Hyperfine Interactions 240 (2019) 97

\bibitem{1976GrZC} J.W.Gruter, B.Jonson, O.B.Nielsen, the ISOLDE Collaboration - CERN-76-13, p.428 (1976)

\bibitem{NNDC185Hg} S.-C. Wu, Nucl. Data Sheets 106 (2005) 619

\bibitem{Briselet2019} R. Briselet, Ch. Theisen, M. Vandebrouck, A. Marchix, M. Airiau, K. Auranen, H. Badran, D. Boilley, T. Calverley, D. Cox, F. D\'echery, F. Defranchi Bisso, A. Drouart, B. Gall, T. Goigoux, T. Grahn, P.T. Greenlees, K. Hauschild, A. Herzan, R.D. Herzberg, U. Jakobsson, R. Julin, S. Juutinen, J. Konki, M. Leino, A. Lightfoot, A. Lopez-Martens, A. Mistry, P. Nieminen, J. Pakarinen, P. Papadakis, J. Partanen, P. Peura, P. Rahkila, J. Rubert, P. Ruotsalainen, M. Sandzelius, J. Saren, C. Scholey, J. Sorri, S. Stolze, B. Sulignano, J. Uusitalo, A. Ward, and M. Zieli\ifmmode \acute{n}\else \'{n}\fi{}ska, Physical Review C 99 (2019) 024614

\bibitem{Hessberger2001} F.P. He{\ss}berger, S. Hofmann, D. Ackermann, V. Ninov, M. Leino, G. M\"{u}nzenberg, S. Saro, A. Lavrentev, A.G. Popeko, A.V. Yeremin, and Ch. Stodel, The European Physical Journal A - Hadrons and Nuclei 12 (2001) 57-67

\bibitem{Rosenbusch2018} M. Rosenbusch, Y. Ito, P. Schury, M. Wada, D. Kaji, K. Morimoto, H. Haba, S. Kimura, H. Koura, M. MacCormick, H. Miyatake, J. Y. Moon, K. Morita, I. Murray, T. Niwase, A. Ozawa, M. Reponen, A. Takamine, T. Tanaka, and H. Wollnik, Physical Review C 97 (2018) 064306

\bibitem{SchuryDoubleCharge} P. Schury, M. Wada, Y. Ito, D. Kaji, H. Haba, Y. Hirayama, S. Kimura, H. Koura, M. MacCormick, H. Miyatake, J. Y. Moon, K. Morimoto, K. Morita, I. Murray, A. Ozawa, M. Rosenbusch, M. Reponen, A. Takamine, T. Tanaka, Y. X. Watanabe and H. Wollnik, Nuclear Instruments and Methods in Physics Research, Section B: Beam Interactions with Materials and Atoms, 407 (2017) 160-165, DOI:10.1016/j.nimb.2017.06.014

\bibitem{NIST} Kramida, A., Ralchenko, Yu., Reader, J., and NIST ASD Team (2019). NIST Atomic Spectra Database (ver. 5.7.1), [Online]. Available: https://physics.nist.gov/asd [2020, March 30]. National Institute of Standards and Technology, Gaithersburg, MD. DOI: https://doi.org/10.18434/T4W30F

\bibitem{AME16} M. Wang, G. Audi, F. G. Kondev, W. J. Huang, S. Naimi, and X. Xu, Chinese Physics C 41 (2017) 030003

\bibitem{Schmidt1984} K.-H. Schmidt, C.-C. Sahm, K. Pielenz, and H.-G. Clerc, Zeitschrift f\"ur Physik A 316 (1984) 19-26

\bibitem{NiwaseThesis} Toshitaka Niwase, \emph{First direct mass measurement of superheavy nuclide via MRTOF mass spectrograph equipped with an $\alpha$-TOF detector}, PhD. Thesis, Department of Physics, Kyushu University, February 2021

\bibitem{ItoSRM} Y. Ito, P. Schury, M. Wada, S. Naimi, T. Sonoda, H. Mita, F. Arai, A. Takamine, K. Okada, A. Ozawa, and H. Wollnik, Physical Review C 88 (2013) 011306(R)

%\bibitem{Wiley-Mclaren} W.C. Wiley and I.H. McLaren, The Review of Scientific Instruments 26 (1955) 1150

%\bibitem{Ringle2009} R. Ringle, G. Bollen, A. Prinke, J. Savory, P. Schury, S. Schwarz, T. Sun, Nuclear Instruments and Methods in Physics Research A 604 (2009) 536-547

\bibitem{Birge} R.T. Birge, Phys. Rev. 40 (1932) 207

\bibitem{Hessberger2001} F.P. He{\ss}berger, S. Hofmann, D. Ackermann, V. Ninov, M. Leino, G. M{\"u}nzenberg, S. Saro, A. Lavrentev, A.G. Popeko, A.V. Yeremin, and Ch. Stodel, Eur. Phys. J. A (2001) 57-67

\bibitem{NUBASE2016}  G. Audi, F. G. Kondev, Wang Meng, Huang W.J., and S. Naimi, Chinese Physics C 41 (2017) 030001

\bibitem{PierreThesis} Pierre Brionnet, \emph{Etude des \'etats isom\`eres des noyaux superlourds :cas des noyaux $^{257}$Db et $^{253}$Lr}, PhD. Thesis, Universit\'e de Strasbourg, September 2017

\bibitem{Hessberger2010} F. P. He{\ss}berger, S. Antalic, D. Ackermann, S. Heinz, S. Hofmann, J. Khuyagbaatar, B. Kindler, I. Kojouharov, B. Lommel, and  R. Mann, Eur. Phys. J. A 43 (2010) 175

\bibitem{DZ1} J. Duflo and A. P. Zuker, Phys. Rev. C 52, R23 (1995)

\bibitem{DZ} J. Mendoza-Temis, J.G. Hirsch, A.P. Zuker, Nuc. Phys. A, Vol. 843, Issues 1-4, 2010, Pgs. 14-36

\bibitem{FRDM12} P. M\"oller, A. Sierk, T. Ichikawa, and H. Sagawa, At. Data Nucl. Data Tables 109 (2016) 1

\bibitem{WS4RBF} N. Wang and M. Liu, Phys. Rev. C 84 (2011) 051303(R)

\bibitem{HFB21} S. Goriely, N. Chamel, J. M. Pearson, Phys. Rev. C 82 (2010) 035804

\bibitem{HFB32} S. Goriely, N. Chamel, and J. M. Pearson, Phys. Rev. C 93 (2016) 034337

\bibitem{WB03} H. Koura, T. Tachibana. Bulletin Physical Society of Japan (BUTSURI), 60 (2005) [Japanese language]

\bibitem{KTUY05} H. Koura, T.Tachibana, M. Uno, M. Yamada, Progr. Theor. Phys. 113 (2005) 305

\end{thebibliography}
\end{document}